\documentclass[12pt,preprint]{aastex}
\usepackage{psfig}
\usepackage{graphicx}

\shorttitle{Neutron Star Binary Systems}
\shortauthors{authors}

\begin{document}

\title{The Probability Distribution of Binary Pulsar Coalescence Rates.
I.~Double Neutron Star Systems in the Galactic Field}

\author{C. Kim\altaffilmark{1}, V. Kalogera\altaffilmark{1} and 
D. R. Lorimer\altaffilmark{2}}

\affil{ $^{1}$ Northwestern University, Dept. of Physics \& Astronomy,
       2145 Sheridan Rd., Evanston, IL 60208\\ $^{2}$ University of
Manchester, Jodrell Bank Observatory, Macclesfield, Cheshire, SK11 9DL, UK.\\
c-kim1@northwestern.edu; vicky@northwestern.edu; drl@jb.man.ac.uk}

\begin{abstract} Estimates of the Galactic coalescence rate (${\cal R}$)
of close binaries with two neutron stars (NS--NS) are known to be
uncertain by large factors (about two orders of magnitude) mainly due to
the small number of systems detected as binary radio pulsars.  We present
an analysis method that allows us to estimate the Galactic NS--NS
coalescence rate using the current observed sample and, importantly, to assign
a statistical significance to these estimates and to calculate the allowed
ranges of values at various confidence levels. The method involves the
simulation of selection effects inherent in all relevant radio pulsar
surveys and a Bayesian statistical analysis for the probability
distribution of the rate. The most likely values for the total Galactic
coalescence rate (${\cal R}_{\rm peak}$) lie in the range $2-60$
Myr$^{-1}$ depending on different pulsar population models. For our
reference model 1, where the most likely estimates of pulsar population
properties are adopted, we obtain ${\cal R}_{\rm tot} = 8_{-5}^{+9}$
Myr$^{-1}$ at a 68\% statistical confidence level. The corresponding range
of expected detection rates of NS--NS inspiral are $3_{-2}^{+4}\times
10^{-3}$ yr$^{-1}$ for the initial LIGO and $18_{-11}^{+21}$ yr$^{-1}$ for
the advanced LIGO. \end{abstract}


\section{INTRODUCTION}
\label{sec:intro}

The detection of the double neutron star (NS--NS) prototype
PSR~B1913+16 as a binary pulsar \cite{ht75} and its orbital decay due
to emission of gravitational waves \cite{tfm79} has inspired a number
of quantitative estimates of the coalescence rate of NS--NS binaries
\cite{cvs79,nps91,ph91,cl95}. In general, the coalescence rate of 
NS--NS binaries can be calculated based on: (a) our theoretical
understanding of their formation (see Belczynski \& Kalogera 2001
\nocite{bk01a} for a review and application of this approach); (b)
the observational properties of the pulsars in the binary systems and
the modeling of pulsar survey selection effects (see \nocite{nar87}
e.g.~Narayan 1987). Interest in these coalescence derives from an
intrinsic motivation of understanding their origin and evolution and
their connections to other NS binaries. However, significant interest
derives from their importance as gravitational-wave sources for the
upcoming ground-based laser interferometers (such as LIGO) and their
possible association with $\gamma$-ray burst events (Popham et al.~1998
\nocite{pop98} and references therein).

The traditional way of calculating the coalescence rate based on
observations involves an estimate of the {\it scale factor}, an
indicator for the number of pulsars in our Galaxy with the same spin
period and luminosity \cite{nar87}. Corrections must then be applied to
these scale factors to account for the faint end of the pulsar
luminosity function, the beamed nature of pulsar emission, and
uncertainties in the assumed spatial distribution.  The estimated
total number in the Galaxy can then be combined with estimates of
their lifetimes to obtain a coalescence rate, ${\cal R}$.  
This method was first
applied by Narayan et al.\ (1991) and Phinney (1991) and other
investigators who followed \cite{cl95,vl96}. Various correction
factors were (or were not) included at various levels of
completeness. Summaries of these earlier studies can be found in
Arzoumanian et al.\ (1999) and Kalogera et al.\ (2001; hereafter
KNST). The latter authors examined all possible
uncertainties in the estimates of the coalescence rate of NS--NS
binaries in detail, and pointed a small-number bias that introduces a
large uncertainty (more than two orders of magnitude) in the
correction factor for the faint-pulsar population that must be applied
to the rate estimate.  They obtained a total NS--NS rate estimate in
the range ${\cal R}=10^{-6} - 5\times 10^{-4}$\,yr$^{-1}$, with the
uncertainty dominated by the small-number bias.  Earlier studies, which
made different assumptions about the pulsar properties (e.g.~luminosity and
spatial distributions and lifetimes), are roughly consistent with
each other (given the large uncertainties). Estimated ranges of values
until now were not associated with statistical significance statements
and an ``all-inclusive'' estimated Galactic coalescence rate lies in
the range $\sim10^{-7} - 10^{-5}$ yr$^{-1}$.

The motivation for this paper is to update the scale factor
calculations using the most recent pulsar surveys, and present a
statistical analysis that allows the calculation of statistical
confidence levels associated with rate estimates.  We consider the two
binaries found in the Galactic disk: PSR~B1913+16 \cite{ht75} and
PSR~B1534+12 \cite{wol91}.  Following the arguments made by Phinney
(1991) and KNST, we do not include the globular cluster system
PSR~B2127+11C \cite{pr91}; this system will be the subject of a later
paper. Radio-pulsar-survey selection effects are taken into account in
the modeling of pulsar population. As described in what follows, the
small-number bias and the effect of a luminosity function are {\it
implicitly included} in our analysis, and therefore a separate
correction factor is not needed.  For each population model of
pulsars, we derive the probability distribution function of the total
Galactic coalescence rate weighted by the two observed binary systems.
In our results we note a number of important correlations between
${\cal R}_{\rm peak}$ and model parameters that are
useful in generalizing the method. We extrapolate the Galactic rate to
cover the detection volume of LIGO and estimate the detection rates of
NS--NS inspiral events for the initial and advanced LIGO.

The plan for the rest of this paper is as follows.  In \S
\ref{sec:method}, we describe our analysis method in a qualitative
way.  Full details of the various pulsar population models and survey
selection effects are then given in \S \ref{sec:models} and \S
\ref{sec:selfx} respectively.  In \S \ref{sec:analysis}, we derive the
probability distribution function for the total Galactic coalescence
rate and calculate the detection rate of LIGO. In \S
\ref{sec:results}, we summarize our results and discuss a number of
intriguing correlations between various physical quantities. Finally,
in \S \ref{sec:discussion}, we discuss the results and compare them
with previous studies.

\section{BASIC ANALYSIS METHOD}
\label{sec:method}

Our basic method is one of ``forward'' analysis. By this we mean that
we do not attempt to invert the observations to obtain the total
number of NS--NS binaries in the Galaxy. Instead, using Monte Carlo
methods, we populate a model galaxy with NS--NS binaries (that match
the spin properties of PSR~B1913+16 and PSR~B1534+12) with pre-set
properties in terms of their spatial distribution and radio pulsar
luminosity function. Details about these ``physical models'' are given
in \S \ref{sec:models}.  

For a given physical model, we produce synthetic populations of
different total numbers of objects ($N_{\rm tot}$).  We then produce a
very large number of Monte Carlo realizations of such pulsar
populations and determine the number of objects ($N_{\rm obs}$) that
are observable by all large-scale pulsar surveys carried out to date
by detailed modeling of the detection thresholds of these
surveys. This analysis utilizes code to take account of observational
selection effects in a self-consistent manner, developed and described
in detail by Lorimer et al.~(1993; hereafter LBDH) which we summarize
in \S \ref{sec:selfx}.  Performing this analysis for many different
Monte Carlo realizations of the physical model allows us to examine
the distribution of $N_{\rm obs}$. We find, as expected and assumed by
other studies, that this distribution closely follows Poisson
statistics, and we determine the best-fit value of the mean of the
Poisson distribution $\lambda$ for each population model and value
of $N_{\rm tot}$.

The calculations described so far are performed separately for each of
PSR~B1913+16 and PSR~B1534+12 so that we obtain separate best-fit
$\lambda$ values for the Poisson distributions. Doing the analysis in
this way allows us to calculate the {\it likelihood of observing just
one example of each pulsar in the real-world sample}.  Given the
Poissonian nature of the distributions this likelihood is simply:
$P(1;\lambda)=\lambda\exp(-\lambda)$. We then calculate this
likelihood for a variety of assumed $N_{\rm tot}$ values for each
physical model.

The probability distribution of the total coalescence rate ${\cal R}_{\rm
tot}$ is derived using the Bayesian analysis and the calculated likelihood
for each pulsar (described in detail in \S \ref{sec:analysis}).  The
derivation of this probability distribution allows us to calculate the
most probable rate as well as determine its ranges of values at
various statistical confidence levels.  Finally, we extrapolate the
Galactic rate to the volume expected to be reached by LIGO and calculate
the detection rate, ${\cal R}_{\rm det}$ (see \S \ref{sec:results}).

\section{MODELS FOR THE GALACTIC PULSAR POPULATION}
\label{sec:models}

Our model pulsar populations are characterized by a Galactocentric
radius ($R$), vertical distance ($Z$) from the Galactic plane and
radio luminosity ($L$).  Assuming that the distributions of each of
these parameters are independent, the combined probability density
function (PDF) of the model pulsar population can be written as:
\begin{equation}
f(R,Z,L) \,\, dR \, dZ \, dL =  \psi_{\rm R}(R) \, 
2\pi \, R \, dR \,\, ~\psi_{\rm Z}(Z) \, dZ \,\, ~\phi(L) \, dL,
\end{equation}
where $\psi_{R}(R)$, $\psi_{Z}(Z)$ and $\phi(L)$, are the
individual PDFs of $R$, $Z$ and $L$,  respectively. In all models
considered, we assume azimuthal symmetry about the Galactic center.

The spatial distribution of pulsars is rather loosely constrained, but we find
that it does not affect the results significantly for a wide range of models.
For the radial and the vertical PDFs, we consider Gaussian and exponential
forms with different values of the radial $R_0$ and the vertical
$Z_0$ scale. In our reference model, we assume a Gaussian PDF for the
radial component and an exponential PDF for the vertical component. Hence, the
combined spatial PDF is given by: 
\begin{equation} f(R,Z) \propto \exp
\left(-\frac{R^2}{2R_0^2}- \frac{\vert Z\vert}{Z_0}\right),
\end{equation} 
We set $R_0=4.0$ kpc and $Z_0=1.5$ kpc as
standard model parameters. Following Narayan et al.~(1991), these and other
values considered reflect the {\it present-day} spatial distribution of the
NS--NS binary population after kinematic evolution in the Galactic
gravitational potential.

Having assigned a position of each pulsar in our model galaxy, for later
computational convenience, we store the positions as Cartesian $x,y,z$
coordinates, where the Galactic center is defined as (0.0,0.0,0.0) kpc and the
position of the Earth is assumed to lie 8.5 kpc from the center along the $x-y$
plane, i.e.~(8.5,0.0,0.0). From these definitions, the distance $d$ to each
pulsar from the Earth can be readily calculated, as well as the apparent
Galactic coordinates $l$ and $b$.

For the luminosity PDF, we follow the results of Cordes \& Chernoff (1997) and
adopt a power-law function of the form \begin{equation} \phi(L) = (p-1) L_{\rm
min}^{\rm p-1} L^{\rm -p}, \end{equation} where $L \geq L_{\rm min}$ and $p>1$.
The cut-off luminosity, $L_{\rm min}$, and the exponent $p$ are the model
parameters.  Cordes \& Chernoff (1997) found $L_{\rm min}=1.1_{-0.5}^{+0.4}$
mJy kpc$^2$ and $p=2.0 \pm 0.2$ at 68\% confidence. We set $p=2.0$ and $L_{\rm
min}=1.0$ mJy kpc$^2$ for our reference model.  Throughout this paper,
luminosities are defined to be at the observing frequency $\nu =400$ MHz.

Having defined the position and luminosity of each pulsar in our model Galaxy,
the final step in defining the model population is to calculate a number of
derived parameters required to characterize the detection of the model pulsars:
dispersion measure (DM), scatter-broadening time ($\tau$) and sky background
temperature ($T_{\rm sky}$).  To calculate DM and $\tau$, we use the software
developed by Taylor \& Cordes (1993) \nocite{tc93} to integrate their model of
the free-electron column density along the line of sight to each pulsar defined
by its model Galactic coordinates $l$ and $b$ out to its distance $d$.  
Frequency scaling of $\tau$ to different survey frequencies is done assuming a
Kolmogorov turbulence spectrum with a spectral index of --4.4\footnote{Although
recent studies suggest a variety of spectral indices for $\tau$ \cite{lkm+01},
the effects of scattering turn out to be negligible in this study since the
detections of NS--NS binaries are limited by luminosity to nearby systems.}.  
Finally, given the model Galactic coordinates of each pulsar, the sky
background noise temperature at 408 MHz ($T_{\rm sky}$) is taken from the
all-sky catalog of \nocite{hssw82} Haslam et al.~(1981). Scaling $T_{\rm sky}$
to other survey frequencies assumes a spectral index of --2.8 \cite{lmop87}.

\section{PULSAR SURVEY SELECTION EFFECTS} \label{sec:selfx}

Having created a model pulsar population with a given spatial and
luminosity distribution, we are now in a position to determine the
fraction of the total population which are actually {\it detectable}
by current large-scale pulsar surveys.  To do this, we need to
calculate, for each model pulsar, the effective signal-to-noise ratio
it would have in each survey and compare this with the corresponding
detection threshold.  Only those pulsars which are nominally above the
threshold count as being detectable. After performing this process on
the entire model pulsar population of size $N_{\rm tot}$, we are left
with a sample of $N_{\rm obs}$ pulsars that are nominally detectable
by the surveys.  Repeating this process many times, we can determine
the probability distribution of $N_{\rm obs}$ which we then use to
constrain the population and coalescence rate of NS--NS binaries. In
this section we discuss our modeling of the various selection effects
which limit pulsar detection.

\subsection{Survey Parameters}

The main factors affecting the signal-to-noise ratio ($\sigma$) of a
pulsar search can be summarized by the following expression
\begin{equation} \sigma \propto \frac{S_{\nu} G}{T} \sqrt{\frac{P
\Delta \nu t}{w_e}}, \end{equation} where $S_{\nu}$ is the apparent
flux density at the survey frequency $\nu$, $G$ is the gain of the
telescope, $T$ is the effective system noise temperature (which
includes a contribution $T_{\rm sky}$ from the sky background
described in the previous section), $P$ is the pulse period, $\Delta
\nu$ is the observing bandwidth, $t$ is the integration time and $w_e$
is the pulse width.  More exact expressions are given in the detailed
description of the survey selection effects in \S 2 of LBDH (in
particular see their eqns.~14--18) which we adopt in this work. In
what follows, we describe the salient points relevant to this study.

For each model pulsar, with known 400-MHz luminosity $L$ and distance $d$, we
calculate the apparent 400-MHz flux density $S_{400} = L/d^2$. Since not all
pulsar surveys are carried out at 400 MHz, we need to scale $S_{400}$ to take
account of the steep radio flux density spectra of pulsars. Using a simple
power law of the form $S_{\nu} \propto \nu^{\alpha}$, where $\alpha$ is the
spectral index, we can calculate the flux $S_{\nu}$ at any frequency $\nu$ as:
\begin{equation}
 S_{\nu} = S_{400} \left( \frac{\nu}{\rm 400 \, MHz} \right)^{\alpha},
\end{equation} 
 Following the results of \nocite{lylg95} Lorimer et al.~(1995), in all
simulations, spectral indices were drawn from a Gaussian PDF with a mean of
--1.6 and standard deviation 0.4.

The telescope gain, system noise, bandwidth and integration time are well-known
parameters for any given survey and the detailed models we use take account of
these.  In addition to the surveys considered by LBDH, we also model surveys
listed by Curran \& Lorimer (1995), and more recent surveys at Green Bank
(Sayer, Nice \& Taylor 1997) and Parkes \cite{lcm+00,mlc+01,ebvb01}. A complete
list of the surveys considered, and the references to the relevant publications
is given in Table 1.

Up to this point in the simulations, the model parameters are identical for
both PSRs B1913+16 and B1534+12. Since we are interested in the individual
contributions each of these systems make to the total Galactic merger rate of
NS--NS binaries similar to these systems, we fix the assumed spin periods $P$
and {\it intrinsic} pulse widths $w$ to the values of each pulsar and perform
separate simulations over all physical models considered. The assumed pulse
widths are 10 ms and 1.5 ms respectively for PSRs B1913+16 and B1534+12. The
{\it effective pulse width} $w_e$ required for the signal-to-noise calculation
must take into account pulse broadening effects due to the interstellar medium
and the response of the observing system. The various contributions are
summarized by the quadrature sum:
 \begin{equation}
  w_e^2 = w^2 + \tau^2 + t_{\rm samp}^2 + t_{\rm DM}^2 + t_{\Delta {\rm DM}}^2,
 \end{equation}
 where $\tau$ is the scatter-broadening timescale calculated from the Taylor \&
Cordes (1993) model, $t_{\rm samp}$ is the data sampling interval in the
observing system, $t_{\rm DM}$ is the dispersive broadening across an
individual frequency channel and $t_{\Delta {\rm DM}}$ is the pulse broadening
due to dedispersion at a slightly incorrect dispersion measure.  All of these
factors are accounted for in our model described in detail in LBDH.

\subsection{Doppler Smearing}

For binary pulsars, we need to take account of the reduction in signal-to-noise
ratio due to the Doppler shift in period during an observation. This was not
considered in LBDH since their analysis was concerned only with isolated
pulsars. For observations of NS--NS binaries, however, where the orbital
periods are of the order of 10 hours or less, the apparent pulse period can
change significantly during a search observation causing the received power to
be spread over a number of frequency bins in the Fourier domain.  As all the
surveys considered in this analysis search for periodicities in the amplitude
spectrum of the Fourier transform of the time series, a signal spread over
several bins can result in a loss of signal-to-noise ratio. To take account of
this effect in our survey simulations, we need to multiply the apparent flux
density of each model pulsar by a ``degradation factor'', $F$.

To calculate the appropriate $F$ values to use, we generate synthetic pulsar
search data containing signals with periods and duty cycles similar to 
PSR~B1913+16 and PSR~B1534+12. 
These data are then passed through a real pulsar search code
which is similar to those in use in the large-scale surveys (see Lorimer et
al.~2000 for details). For each of the two pulsars, we first generate a
control time series in which the signal has a constant period and find the
resulting signal-to-noise ratio, $\sigma_{\rm control}$, reported by the search
code. We then generate a time series in which the pulses have identical
intensity but are modulated in period according to the appropriate orbital
parameters of each binary pulsar. From the resulting search signal-to-noise
ratio, $\sigma_{\rm binary}$, the degradation factor $F=\sigma_{\rm
binary}/\sigma_{\rm control}$.  Significant degradation occurs, therefore, when
$F \ll 1$.  Since accumulated Doppler shift, and therefore $F$, is a strong
function of the orbital phase at the start of a given observation, for both
binary systems, we calculate the mean value of $F$ for a variety of starting
orbital phases appropriately weighted by the time spent in that particular part
of the orbit.

A similar analysis was made by Camilo et al.~(2000) for the millisecond pulsars
in 47~Tucanae. In this paper, where we are interested in the degradation as a
function of integration time, we generate time series with a variety of
lengths between 1 minute and 1 hour using sampling intervals similar to those
of the actual surveys listed in Table 1. The results are
summarized in Figure~\ref{fig:fvst}, where we plot average $F$ versus integration
time for both sets of orbital parameters.  As expected, surveys with the
longest integration times are most affected by Doppler smearing. For the Parkes
Multibeam survey \cite{lcm+00,mlc+01}, which has an integration time of 35 min,
mean values of $F$ are 0.7 and 0.3 for PSR~B1913+16 and PSR~B1534+12 respectively\footnote{In order to improve on the sensitivity to binary pulsars,
the Parkes Multibeam survey data are now being reprocessed using various
algorithms designed to account for binary motion during the integration time
(Faulkner et al.~2002)}. The greater degradation for PSR~B1534+12 is due to its
mildly eccentric orbit ($e \sim 0.3$ versus 0.6 for PSR~B1913+16) which results in
a much more persistent change in apparent pulse period when averaged over the
entire orbit.  For the Jodrell Bank and Swinburne surveys \cite{nll+95,ebvb01},
which both have integration times of order 5 min, we find $F \sim 0.9$ for both
systems.  For all other surveys, which have significantly shorter integration
times, no significant degradation is seen, and we take $F=1$.

\section{STATISTICAL ANALYSIS}
\label{sec:analysis}

In this section we describe in detail the derivation of the probability
distribution of the Galactic coalescence rate ${\cal R}$. 
The analysis method makes
use of Bayesian statistics and takes into account the rate contributions of
both observed NS--NS binaries. At the end of the section we derive the
associated detection rates for LIGO.

\subsection{The Rate Probability Distribution for Each Observed NS--NS Binary }

As already mentioned in \S\,\ref{sec:method}, for each of the two observed
NS--NS binaries (PSR~B1913+16 or PSR~B1534+12) we generate pulsar populations
in physical and radio luminosity space with pulse periods and widths fixed to
the observed ones and with different absolute normalizations, i.e., total
number $N_{\rm tot}$ of pulsars in the Galaxy.  We generate large numbers of
``observed'' pulsar samples by modeling the pulsar survey selection effects
(see \S\,\ref{sec:selfx}) and applying them on these model populations of
PSR~B1913+16--like and PSR~B1534+12--like pulsars separately (see
\S\,\ref{sec:models}). For a fixed value of $N_{\rm tot}$, we use these
``observed'' samples to calculate the distribution of the number of objects in
the samples. One might expect that the number of observed pulsars in a sample
$N_{\rm obs}$ follows very closely a Poisson distribution:
 \begin{equation}
 P( N_{\rm obs}; \lambda )~=~\frac{\lambda^{N_{\rm obs}}~e^{-\lambda}}{N_{\rm
obs}!},
 \end{equation}
where, by definition, $\lambda\equiv <N_{\rm obs}>$.  We confirm our expectation by obtaining
excellent formal fits to the Monte Carlo data using such a distribution and
calculating the best-fitting value of $\lambda$. We vary $N_{\rm tot}$ in the
range $10-10^{4}$ and find that $\lambda$ is linearly correlated with
$N_{\rm tot}$:
 \begin{equation}
 \label{eq:lamb}
 \lambda = \alpha N_{\rm tot},
 \end{equation}
 where $\alpha$ is a constant that depends on the properties (space and
luminosity distributions and pulse period and width) of the Galactic pulsar
population.  Examples of the Poisson fits and the linear correlations are shown
in Figures \ref{fig:poisson} and \ref{fig:lambda}, respectively, for our
reference model 1 (see Table 2 and \S\,6).

The main step in deriving a rate probability distribution for each of the
observed systems is to first derive the probability distribution of the total
number of pulsars like the observed ones in the Galaxy. We obtain the latter
by applying Bayes' theorem:
 \begin{equation}
 P(H|DX)=P(H|X) \frac{P(D|HX)}{P(D|X)},
 \end{equation}
 where $P(H|DX)$ is the probability of a model hypothesis $H$ given data $D$
and model priors $X$, $P(D|HX)$ is the likelihood of the data given a model
hypothesis and priors, $P(H|X)$ is the probability of a model hypothesis in the
absence of any data information, and $P(D|X)$ is the model prior probability,
which acts as a normalization constant.

In the present work, we make following identifications:
 \begin{math}
 \\
 D: {\rm is~the~real~observed~sample} \\
 H: {\rm is}~\lambda~proportional~to~ N_{\rm tot} \\
 X: {\rm is~the~population~model~(space~and~luminosity~distributions~and~pulse~period~and~width)}
 \end{math}
 With these identifications, $P(D|HX)$ is the likelihood of the real observed
sample (one ``PSR~B1913+16--like'' and one ``PSR~B1534+12--like'' pulsar) and is
obtained by the best-fitting Poisson distribution:
 \begin{equation}
 P(D|HX)~=~P\left(1; \lambda(N_{\rm tot}), X\right)~=~\lambda(N_{\rm
tot})~e^{-\lambda(N_{\rm tot})}.
 \end{equation}
 In the absence of any data information, the absolute normalization of the
model population, i.e., the total pulsar number $N_{\rm tot}$ and hence
$\lambda$ is expected to be {\em independent} of the shape of the population
distributions and properties represented by $X$. Therefore, the probability of
$\lambda$ given a set of assumptions $X$ for the model Galactic population is
expected to be flat:
 \begin{equation}
 P(H|X)~=~P(\lambda(N_{\rm tot}) | X)~=~{\rm constant},
 \end{equation}
 and is essentially absorbed by the model prior probability $P(D|X)$ as a
normalization constant. The probability distribution of $\lambda$ then is
given by:
\begin{equation}
 P(\lambda | DX)~=~\frac{P(\lambda|X)}{P(D|X)}~P(1; \lambda, X)~=~{\rm
constant}~\times~P(1; \lambda, X).
 \end{equation}
 We impose the normalization constraint:
 \begin{math}
 \int^\infty_0 P(\lambda | DX) d\lambda=1
 \end{math}
 and find that $P(H|X)/P(D|X)=1$ and
 \begin{equation}
 P(\lambda | DX)~=~P(1; \lambda, X)~=~\lambda~e^{-\lambda}.
 \end{equation}
 Note that based on the above expression, the maximum value of $P(\lambda)$
equal to $e^{-1}$ always occurs at $\lambda=1$ or at $N_{\rm tot}=\alpha^{-1}$
(see eq.(\ref{eq:lamb})). It is straightforward to calculate $P(N_{\rm tot})$:
 \begin{eqnarray}
 P(N_{\rm tot}) & = & P(\lambda) \Bigl|{ {d\lambda} \over {dN_{\rm tot}} } \Bigr| \nonumber \\ 
& = & \alpha^{2} N_{\rm tot} e^{-\alpha N_{\rm tot}}
 \end{eqnarray}

For a given total number of pulsars in the Galaxy, we can calculate their rate
using estimates of the associated pulsar beaming correction factor $f_{\rm b}$
and lifetime $\tau_{\rm life}$:
 \begin{equation}
 \label{eq:rco}
 {\cal R}~=~\frac{N_{\rm tot}}{\tau_{\rm life}} f_{\rm b}.
 \end{equation}
 Note that this estimate is equivalent to previous studies (KNST and
references therein) where the concept of the {\em scale factor} is used instead
of our calculated $N_{\rm tot}$. We can write equivalently for our calculation:
 \begin{equation}
 \frac{N_{\rm tot}}{N_{\rm obs}}~={\frac {\int \int_{\rm V_{G}} f(R,Z,L)
dVdL} {\int \int_{\rm V_{D}} f(R,Z,L)  dVdL}},
 \end{equation}
 where $f(R,Z,L)$ is the probability distribution of the pulsar population
(\S\,\ref{sec:models}), $V_{\rm G}$ and $V_{\rm D}$ are the Galactic volume and
the detection volume (for pulsars with pulse period and width similar to each
of the two observed pulsars), respectively. Note that we do not fix the
luminosity of the pulsar population to the observed values, but instead, we
estimate $N_{\rm tot}$ for a distribution of radio luminosities. Since we
consider separately PSR~B1913+16--like and PSR~B1534+12--like populations,
$N_{\rm obs}$= 1.

For pulsar beaming fractions, we adopt the estimates obtained by KNST:
5.72 for PSR~B1913+16 and 6.45 for PSR~B1534+12. For the lifetime
estimates we also follow KNST. Our adopted values for the pulsar lifetimes
are $3.65 \times 10^{8}$\,yr for PSR~B1913+16 and $2.9\times 10^{9}$\,yr
for PSR~B1534+12.
        
Using eq. (\ref{eq:rco}) we calculate $P({\cal R})$ for each of the two observed
pulsars:
 \begin{eqnarray}
 P({\cal R}) & = & P(N_{\rm tot}) \Bigl|{dN_{\rm tot} \over d{\cal R}}\Bigr| \nonumber \\ 
& = & \Bigl({\frac{\alpha \tau_{\rm life}}{f_{\rm b}}} \Bigr)^{2} {\cal R}~
e^{-\bigl({\frac{\alpha \tau_{\rm life}}{f_{\rm b}}} \bigr) {\cal R}}.
 \end{eqnarray}

\subsection{The Total Galactic Coalescence Rate}

Once the probability distributions of the rate contributions of the two
observed pulsars are calculated, we can obtain the distribution functions of
the total coalescence rate ${\cal R}_{\rm tot}$. We define the following two
coefficients for each observed system:
 \begin{equation}
 A \equiv {\biggl({ \alpha \tau_{\rm life} \over f_{\rm b} }\biggr)}_{\rm
1913}~~{\rm and }~~ B \equiv {\biggl({ \alpha \tau_{\rm life} \over f_{\rm b}
}\biggr)}_{\rm 1534}
 \end{equation}
 and rewrite the coalescence rate for each binary system:
 \begin{equation}
 P_{\rm 1913}({\cal R})= A^{2}{\cal R}e^{-A{\cal R}} ~~{\rm and }~~ P_{\rm 1534}({\cal R})=B^{2}{\cal R}e^{-B{\cal R}}.
 \end{equation}
 One can confirm that each distribution function satisfies the normalization
condition
 \begin{equation}
 \int_0^\infty P({\cal R}) d{\cal R} = 1.
 \end{equation}
 We then define two new variables: ${\cal R}_{\rm +} \equiv {\cal R}_{\rm 1913} + {\cal R}_{\rm
1534}$ and ${\cal R}_{\rm -} \equiv {\cal R}_{\rm 1913} - {\cal R}_{\rm 1534}$. Since
${\cal R}_{\rm 1913} \geq 0$ and ${\cal R}_{\rm 1534} \geq 0$, we have
$-{\cal R}_{\rm +} \leq {\cal R}_{\rm -} \leq {\cal R}_{\rm +}$. For convenience, we rename
${\cal R}_{\rm 1913}={\cal R}_{\rm 1}$ and ${\cal R}_{\rm 1534}={\cal R}_{\rm 2}$ and perform a
two-dimensional probability distribution transformation:
 \begin{equation}
 P({\cal R}_{\rm +}, {\cal R}_{\rm -}) = P({\cal R}_{\rm 1},{\cal R}_{\rm 2}) \left \vert \matrix {
{{d{\cal R}_{\rm 1}} \over {d{\cal R}_{\rm +}} } & {{d{\cal R}_{\rm 2}} \over {d{\cal R}_{\rm -}} } \cr
{d{\cal R}_{\rm 1} \over d{\cal R}_{\rm -} } & {d{\cal R}_{\rm 2} \over d{\cal R}_{\rm +}} \cr } \right
\vert = {1 \over 2} P({\cal R}_{\rm 1},{\cal R}_{\rm 2}).
 \end{equation}
 The probability distribution of the total rate ${\cal R}_{\rm tot}\equiv {\cal R}_{\rm +}$
is obtained after integrating $P({\cal R}_{\rm +}, {\cal R}_{\rm -})$ over ${\cal R}_{\rm -}$:
\begin{equation}
 P({\cal R}_{\rm +}) = \int_{\rm {\cal R}_{\rm -}} P({\cal R}_{\rm +},{\cal R}_{\rm -}) \, d{\cal R}_{\rm -} =
 {1 \over 2} \int_{\rm {\cal R}_{\rm -}} P({\cal R}_{\rm 1}, {\cal R}_{\rm 2}) \, d{\cal R}_{\rm -}. 
 \end{equation}
 Since the probability distributions for the rate contributions of each of the
two observed pulsars are independent of each other, their two-dimensional
distribution is simply the product of the two.
 \begin{equation}
 P({\cal R}_{\rm 1},{\cal R}_{\rm 2}) = P({\cal R}_{\rm 1})P({\cal R}_{\rm 2}).
 \end{equation}
 After we replace all variables to ${\cal R}_{\rm +}$ and ${\cal R}_{\rm -}$, we have
 \begin{eqnarray} \label{eq:Prtot}
 P({\cal R}_{\rm +} \equiv {\cal R}_{\rm tot}) & = & {A^{2}B^{2} \over 2^{3}} e^{-\bigl({A+B \over 2}
\bigr){\cal R}_{\rm +}} {\Bigl[ {\cal R}_{\rm +}^{2} \int_{\rm -{\cal R}_{\rm +}}^{\rm +{\cal R}_{\rm +}}
d{\cal R}_{\rm -} e^{\bigl( {B-A \over 2 }\bigr){\cal R}_{\rm -}} - \int_{\rm -{\cal R}_{\rm
+}}^{\rm +{\cal R}_{\rm +}} d{\cal R}_{\rm -} {\cal R}_{-}^{2} e^{\bigl({B-A \over 2 }\bigr){\cal R}_{\rm
-}} \Bigr]} \nonumber \\
 & = & {\Bigl( {AB \over B-A}\Bigr)^{2}} \Bigl[ {{\cal R}_{\rm +}} {\bigl({
e^{-A{\cal R}_{\rm +}} + e^{-B{\cal R}_{\rm +}} }\bigr)} - {\Bigl( {2 \over B-A} \Bigr)}
{\bigl({ e^{-A{\cal R}_{\rm +}} - e^{-B{\cal R}_{\rm +}} }\bigr)} \Bigr].
 \end{eqnarray}
We have confirmed that the above function satisfies the normalization
$\int^\infty_0 P({\cal R}_{\rm tot}) d{\cal R}_{\rm tot}~=~1$.

Having calculated the probability distribution of the Galactic coalescence
rate, we can take one step further and also calculate ranges of values for the
rate ${\cal R}_{\rm tot}$ at various confidence levels (CL).
The lower (${\cal R}_{\rm a}$) and upper (${\cal R}_{\rm b}$) limits to these ranges
are calculated using:
 \begin{equation}
 \int_{{\cal R}_{\rm a}}^{{\cal R}_{\rm b}} P({\cal R}_{\rm tot}) d{\cal R}_{\rm tot} = {\rm CL}
 \end{equation}
and
\begin{equation}
 P({\cal R}_{\rm a})=P({\cal R}_{\rm b}).
 \end{equation}
In all our results, we quote coalescence rates for 68\%, 95\% and 99\% CL.

\subsection{The Detection Rate for LIGO} \label{sec:ligo}

NS--NS binary systems are expected to emit strong gravitational waves during
their inspiral phase, the late stages of which may be detected by the
ground-based gravitational wave detectors. In this paper we estimate expected
detection rates for LIGO (initial and advanced) using the derived probability
distribution of the Galactic rate and its values at the maximum of the
distribution, for different physical models of pulsar properties.  To calculate
the detection rate, we need to extrapolate the Galactic rate to the volume
detectable by LIGO. We use the ratio between the B-band luminosity density of
the Universe and the B-band luminosity of our Galaxy as the scaling factor
(Phinney 1991; KNST). This is based on the assumptions that (i) the B-band
luminosity (corrected for dust absorption) correlates with the star--formation
rate in the nearby universe and hence the coalescence rate, (ii) the B-band
luminosity density is constant in the nearby universe.  The detection rate,
${\cal R}_{\rm det}$, is calculated by the following equation:

 \begin{equation}
 {\cal R}_{\rm det} = \epsilon {\cal R}_{\rm tot} V_{\rm det},
 \end{equation}
 where $\epsilon$ is the scaling factor assumed to be $\simeq
10^{-2}$\,Mpc$^{-3}$ (for details see KNST). $V_{\rm det}$ is the
detection volume defined as a sphere with a radius equals to the maximum detection distance D$_{\rm max}$ for the initial ($\simeq 20$\,Mpc) and advanced LIGO
($\simeq 350$\,Mpc) \cite{f01}.

\section{RESULTS} \label{sec:results}

We have calculated the probability distribution of the Galactic coalescence
rate ${\cal R}_{\rm tot}$, its most likely value ${\cal R}_{\rm peak}$ and ranges at
different statistical confidence levels, and the most likely expected detection
rates for the initial and advance LIGO for a large number of model pulsar
population properties (see Table 2). We have chosen one of them to be our {\em
reference} model based on the statistical analyses and results presented by
 Cordes \& Chernoff (1997).

For our reference model ($L_{\rm min}=1.0$\,mJy \,kpc$^2$, $p=2.0$,
$R_{\rm o}=4.0$\,kpc, and $Z_0=1.5$\,kpc), we find the most likely value
of $N_{\rm tot}$, to be $\simeq 390$ pulsars for the ``PSR~B1913+16-like''
population, and $\simeq 350$ pulsars for the ``PSR~B1534+12-like''
population. Using eq. (\ref{eq:Prtot}), we evaluate the total Galactic
coalescence rate of NS--NS binaries for this reference case (model 1 in
Table 2). The most likely value of the coalescence rate is $\simeq
8$\,Myr$^{-1}$ and the ranges at different statistical confidence levels
are: $\sim 3-20$\ Myr$^{-1}$ at 68\%, $\sim 1-30$\ Myr$^{-1}$ at 95\%, and
$\sim 0.7-40$\ Myr$^{-1}$ at 99\%.

In Figure \ref{fig:pdf}, $P({\cal R}_{\rm tot})$ along with $P({\cal R}_{\rm 1913})$ and
$P({\cal R}_{\rm 1534})$ are shown for the reference model. It is evident that the
total rate distribution is dominated by that of PSR~B1913+16. At first this
appears to be in contradiction to most other studies of the NS--NS coalescence
rate (Narayan et al. 1991; Phinney 1991; Curran \& Lorimer 1995; van den Heuvel \& Lorimer 1996; Stairs et al. 1998; Arzoumanian et al. 1999; KNST). However, it turns out that
this difference is due to the fact that earlier studies restricted the scale
factor calculation to the actual observed pulsar luminosity and the rate
estimates were dominated by the low luminosity of, and hence large scale factor
for, PSR~B1534+12. Any corrections to the rate estimate which take into account the range of pulsar luminosities were applied as subsequent upward corrections. However,
this effect is eliminated in our case because we calculate the two rate
contributions having relaxed the luminosity constraint and instead having
allowed for a range in luminosity for the pulsars. In this case any differences
in the two separate rate contributions depend {\em only} on differences in
pulse periods, and widths. Given that the latter are rather small, it makes
sense that, for example, the most likely values of $N_{\rm tot}$ for the two pulsars come out
to be very similar (e.g. $\simeq 390$ and $\simeq 350$, for PSR~B1913+16 and
PSR~B1534+12, respectively, in model 1). 
Consequently any difference in the rate
contributions from the two populations is due to the difference in lifetimes
(about a factor of 10) for the two observed pulsars (note that the two do not
only have similar $N_{\rm tot}$ estimates, but also similar beaming correction
factors).  Since the lifetime estimate for PSR~B1913+16 is much smaller, the
total rate distribution is dominated by its contribution.

In Table 2 we list the population parameters and results of Galactic
coalescence and LIGO detection rates for a large number of models (but
still a subset of the models we have investigated in detail). Model 1 is
our reference model defined above and the rest of the models are used to
explore the sensitivity of our results on the pulsar population properties
in luminosity and space distributions. The variation of the luminosity
function parameters is within ranges that correspond to 68\% confidence
levels from the statistical analysis of Cordes \& Chernoff (1997). The
variation of the space scale lengths $R_0$ and $Z_0$ extends to values
that are not favored by our current understanding, only to allow us to
examine the presence of any correlations.  
Scrutiny of the results presented in Table 2 reveals that rate estimates
are modestly sensitive (and in some cases, e.g., $Z_0$, insensitive) to
most of these variations, except in cases of very low values of $L_{\rm
min}$, down to 0.3 mJy kpc$^{2}$, and unphysically low values of $R_0$,
down to $2-3$\,kpc. Assumptions about the shape of the space distribution
(exponential or Gaussian) are not important. The most important model
parameter seems to be the slope of the luminosity function.  The most
likely values for the rates are found in the range $\simeq 3-22$\,Myr$^{-1}$ 
for all models with luminosity-function
parameters consistent with a 68\% confidence level 
estimated by Cordes \& Chernoff (1997) and $R_0$ in the range $4-8$\,kpc. 
Within a given model the range of estimated values at 68\% confidence
level is typically broad by a factor of 5, while at the 95\% level the
ranges broaden by another factor of about 5, reaching an uncertainty of
$\sim 25$ for the rate estimates.

In terms of qualitative variations, it is clear that models with
increasing fraction of low-luminosity or very distant pulsars lead to
increasing rate estimates, as expected. A demonstration of this kind
of dependence is shown in Figures \ref{fig:corr1} and
\ref{fig:corr2}. We have found that there is a strong correlation
between the peak value of the total Galactic coalescence rate ${\cal
R}_{\rm peak}$ estimated by eq. (\ref{eq:Prtot}) and the cut-off
luminosity $L_{\rm min}$ and the power index p. As seen in these
Figures ${\cal R}_{\rm peak}$ increases rapidly with decreasing
$L_{\rm min}$ (Fig. \ref{fig:corr1}) or with increasing p
(Fig. \ref{fig:corr2}).  Scale lengths of the spatial distribution
(either $R_0$ or $Z_0$) do not show a correlation with
${\cal R}_{\rm peak}$ as strong as that of the luminosity-function
parameters.  In Figure \ref{fig:corr3} the most likely rate estimate
is plotted as a function of the radial scale length $R_0$. It is
evident that the rate is relatively insensitive to $R_0$
variations unless $R_0$ becomes very small ($\lesssim
3$\,kpc). Small values of the radial scale for the Galactic
distribution imply a large fraction of very distant (and hence
undetectable) pulsars in the Galaxy and lead to a rapid increase of
the estimated coalescence rate. However, such small $R_0$ values
are not consistent with our current understanding of stellar
populations in the Galaxy \cite{lmt85}.

\section{DISCUSSION}
 \label{sec:discussion}

In this paper we present a new method of estimating the total number of pulsars
in our Galaxy and we apply it to the calculation of the coalescence rate of
double neutron star systems in the Galactic field. The method implicitly takes
into account the small number of pulsars in the observed double-neutron-star
sample as well as their distribution in luminosity and space in the Galaxy. The
modeling of pulsar survey selection effects is formulated in a ``forward'' way,
by populating the Galaxy with model pulsar populations and calculating the
likelihood of the real observed sample. This is in contrast to the ``inverse''
way of the calculation of scale factors used in previous studies. The
formulation presented here allows us to: (a) calculate the probability
distribution of coalescence rates; (b) assign statistical significance to
these estimates; and (c) quantify the uncertainties associated with them.  

As originally shown by KNST, the most important uncertainties
originate from the combination of a small-number observed pulsar and a pulsar
population dominated by faint objects. The probability distribution covers
more than 2 orders of magnitude in
agreement with the uncertainties in excess of two orders of magnitudes asserted
by KNST. However, for the reference model, even at high statistical confidence level (99\%) the uncertainty is reduced to a factor of $\sim 60$. At confidence levels of 95\% and 68\%, the 
uncertainty is further reduced to just $\sim 25$ and $\sim 5$, respectively. 
We use our results to estimate the expected detection rates for ground-based 
interferometers, such as LIGO. The most likely values are found in the range 
$\sim (1 - 30) \times 10^{-3}$\,yr$^{-1}$ and $\sim 4 - 140$\,yr$^{-1}$, for 
the initial and advanced LIGO. 

The statistical method developed here can be further extended to account for
distributions of pulsar populations in pulse periods, widths, and orbital
periods. Most importantly the method can be applied to any type of pulsar
population with appropriate modifications of the modeling of survey selection
effects.  Currently we are working on assessing the contribution of double
neutron stars formed in globular clusters as well as the formation rate of
binary pulsars with white dwarf companions that are important for
gravitational-wave detection by LISA, the space-based interferometer planned by
NASA and ESA for the end of this decade.

\acknowledgments
 This work is partially supported by NSF grant PHY-0121420 to VK.  DRL is a
University Research Fellow funded by the Royal Society. DRL is also grateful
for the hospitality and support of the Theoretical Astrophysics Group at
Northwestern U.  VK would also like to acknowledge partial support by the Clay
Fellowship at the Smithsonian Astrophysical Observatory where this work was
initiated.

\appendix

\clearpage
 \begin{deluxetable}{rrrrrrrrc}
 \label{tab:surveys}
 \tablewidth{34pc}
 \tabletypesize{\footnotesize}
 \tablecaption{Simulated Pulsar Surveys}
 \tablehead{
 \colhead{Telescope} &
 \colhead{Year} &
 \colhead{$\nu^{1}$} &
 \colhead{$\Delta \nu^{2}$} &
 \colhead{$t_{\rm obs}^{3}$} &
 \colhead{$t_{\rm samp}^{4}$} &
 \colhead{$S_{\rm min}^{5}$} &
 \colhead{Detected$^{6}$} &
 \colhead{Refs$^{7}$} 
}
 \startdata
 Lovell 76 m  & 1972 & 408 & 4 & 660 &40& 10& 51/31    & 1,2 \\
 Arecibo 305 m& 1974 & 430 & 8 & 137 &17& 1 & 50/40    & 3,4\\
 Molonglo     & 1977 & 408 & 4 & 45  &20& 10& 224/155  & 5\\
 Green Bank 300'&1977& 400 &16 &138  &17&10 & 50/23    & 6,7\\
 Green Bank 300'&1982& 390 & 16&138  &17& 2 & 83/34    & 8\\
 Green Bank 300'&1983& 390 & 8 &132  &2& 5 & 87/20    & 9\\
 Lovell 76 m  & 1983 &1400 & 40&524  &2& 1 & 61/40    & 10\\
 Arecibo 305 m& 1984 & 430 & 1 &40   &0.3& 3 & 24/5     & 9\\
 Molonglo     & 1985 & 843 & 3 &132  &0.5& 8 & 10/1     & 11\\
 Arecibo 305 m& 1987 & 430 & 10&68   &0.5& 1 & 61/24    & 12\\
 Parkes  64 m & 1988 &1520 &320&150  &0.3& 1 & 100/46   & 13\\
 Arecibo 305 m& 1990 & 430 & 10&40   &0.5& 2 & 2/2      & 14\\
 Parkes  64 m & 1992 & 430 & 32&168  &0.3& 3 & 298/101  & 15,16\\
 Arecibo 305 m& 1993 & 430 & 10&40   &0.5& 1 & 56/90  & 17--20\\
 Lovell 76 m  & 1994 & 411 & 8 &315  &0.3& 5 & 5/1      & 21 \\
 Green Bank 140'&1995&370  & 40&134  &0.3& 8 & 84/8     & 22\\
 Parkes 64 m  & 1998 &1374 &288&265  &0.1&0.5& 69/170   & 23\\
 Parkes 64 m  & 1998 &1374 &288&2100 &0.3&0.2& $\sim$900/600 & 24,25\\

 \enddata
\tablenotetext{1}{Center frequency in MHz.}
\tablenotetext{2}{Bandwidth in MHz.}
\tablenotetext{3}{Integration time in seconds.}
\tablenotetext{4}{Sampling time in milliseconds.}
\tablenotetext{5}{Sensitivity limit in mJy at the survey frequency for
long-period pulsars (calculated for each trial in the simulations).}
\tablenotetext{6}{Total number of detections and new pulsars}
\tablenotetext{7}{ References:
 (1,2) Davies, Lyne \& Seiradakis (1972,3);
 (3,4) Hulse \& Taylor (1974,5);
 (5) Manchester et al.~(1978);
 (6,7) Damashek et al.~(1978,1982);
 (8) Dewey et al.~(1985);
 (9) Stokes et al.~(1986);
 (10) Clifton et al.~(1992);
 (11) D'Amico et al.~(1988);
 (12) Nice, Taylor \& Fruchter (1995);
 (13) Johnston et al.~(1992);
 (14) Wolszczan (1991);
 (15) Manchester et al.~(1996);
 (16) Lyne et al.~(1998);
 (17) Ray et al.~(1996);
 (18) Camilo et al.~(1996);
 (19) Foster et al.~(1995);
 (20) Lundgren, Zepka \& Cordes (1995);
 (21) Nicastro et al.~(1995);
 (22) Sayer, Nice \& Taylor (1997);
 (23) Edwards et al.~(2001);
 (24) Lyne et al.~(2000);
 (25) Manchester et al.~(2002)
}

 \end{deluxetable}
 \nocite{dls72,dls73}
 \nocite{ht74,ht75}
 \nocite{mlt+78}
 \nocite{dth78,dbtb82}
 \nocite{dtws85}
 \nocite{sstd86}
 \nocite{clj+92}
 \nocite{dmd+88}
 \nocite{nft95}
 \nocite{jlm+92}
 \nocite{wol91}
 \nocite{mld+96}
 \nocite{lml+98}
 \nocite{rpj+96}
 \nocite{cnst96}
 \nocite{fcwa95}
 \nocite{lzc95}
 \nocite{nll+95}
 \nocite{snt97}
 \nocite{ebvb01}
 \nocite{lcm+00}
 \nocite{mlc+01}

\clearpage
 
 \begin{deluxetable}{lclllrllrlrl}
 \label{tab:results}

 \tablewidth{0pc}
 \tabletypesize{\small}
 \tablecaption{Model Parameters and Estimates for ${\cal R}_{\rm tot}$ and ${\cal R}_{\rm
det}$ at Various Statistical Confidence Levels for Different Population
Models}
 \tablehead{
 \colhead{Model$^{1}$} & \multicolumn{4}{c}{Parameters}
 & \multicolumn{3}{c}{${\cal R}_{\rm tot}$ (Myr$^{-1}$)}
 & \multicolumn{4}{c}{${\cal R}_{\rm det}$ of LIGO (yr$^{-1}$)}\\
 
\cline{1-12}
 
\colhead{} & \colhead{$L_{\rm min}^{2}$} & \colhead{p$^{3}$} &
\colhead{$R_0^{4}$} & \colhead{$Z_0^{5}$} &\colhead{peak$^{7}$}
&\colhead{68\%$^{8}$} &\colhead{95\%$^{8}$} & \multicolumn{2}{c}{initial
($\times 10^{-3}$)} &\multicolumn{2}{c}{advanced} \\
 
\cline{9-10} \cline{11-12}
 
\colhead{} & \colhead{(mJy kpc$^{2}$)} & \colhead{} & \colhead{(kpc)} &
\colhead{(kpc)} &\colhead{} &\colhead{} &\colhead{} & \colhead{peak$^{7}$} &
\colhead{68\%$^{8}$} &\colhead{peak$^{7}$} &\colhead{68\%$^{8}$}
}
 
\startdata
 1 & 1.0 & 2.0 & 4.0 (G$^{6}$) & 1.5 (E$^{6}$) & 8.0 &$_{-4.7}^{+9.3}$ & $_{-6.7}^{+23.3}$ & 3.3 &$_{-2.0}^{+3.9}$ & 17.9 &$_{-10.6}^{+21.0}$ \\
 2 & 1.0 & 2.0 & 4.0 (G) & 0.5 (E) & 7.1 &$_{-4.2}^{+8.3}$ & $_{-5.9}^{+20.8}$ & 3.0 &$_{-1.7}^{+3.5}$ & 15.9 &$_{-9.4}^{+18.7}$ \\
 3 & 1.0 & 2.0 & 4.0 (G) & 2.0 (E) & 8.4 &$_{-5.0}^{+9.9}$ & $_{-7.1}^{+24.7}$ & 3.5 &$_{-2.1}^{+4.1}$  & 19.0  &$_{-11.2}^{+22.2}$ \\
 4 & 1.0 & 2.0 & 4.0 (E) & 1.5 (E) & 8.7 & $_{-5.1}^{+10.2}$ & $_{-7.3}^{+25.6}$ & 3.6 &$_{-2.2}^{+4.3}$ & 19.5 &$_{-11.6}^{+23.0}$ \\
 5 & 1.0 & 2.0 & 4.0 (G) & 1.5 (G) &  7.9 &$_{-4.6}^{+9.2}$ & $_{-6.6}^{+23.0}$
& 3.3 &$_{-1.9}^{+3.9}$ & 17.7 &$_{-10.}^{+20.7}$\\
 \\
 6 & 0.3 & 2.0 & 4.0 (G) & 1.5 (E) & 26.9 &$_{-16.1}^{+32.0}$ & $_{-22.7}^{+80.3}$& 11.3 &$_{-6.7}^{+13.4}$ & 60.5 &$_{-36.2}^{+72.1}$ \\
 7 & 0.7 & 2.0 & 4.0 (G) & 1.5 (E) & 11.5 &$_{-6.8}^{+13.5}$ & $_{-9.7}^{+33.9}$ &  4.8 &$_{-2.9}^{+5.7}$ & 25.9 &$_{-15.3}^{+30.5}$\\
 8 & 1.5 & 2.0 & 4.0 (G) & 1.5 (E) &  5.5 &$_{-3.2}^{+6.3}$ & $_{-4.6}^{+15.9}$
 & 2.3 &$_{-1.3}^{+2.7}$ & 12.3 & $_{-7.2}^{+14.3}$\\
 9 & 3.0 & 2.0 & 4.0 (G) & 1.5 (E) & 2.9 &$_{-1.7}^{+3.3}$ & $_{-2.4}^{+8.3}$
& 1.2 &$_{-0.7}^{+1.4}$ & 6.4 &$_{-3.8}^{+7.4}$\\
\\
10 & 0.3 & 1.8 & 4.0 (G) & 1.5 (E) & 9.4 & $_{-5.5}^{+10.8}$ & $_{-7.9}^{+27.1}$  &  3.9 &$_{-2.3}^{+4.5}$ & 21.2 &$_{-12.4}^{+24.3}$ \\
11 & 0.7 & 1.8 & 4.0 (G) & 1.5 (E) & 4.8  & $_{-2.8}^{+5.4}$ & $_{-4.0}^{+13.5}$  &  2.0 &$_{-1.2}^{+2.3}$ & 10.7 &$_{-6.2}^{+12.2}$ \\
12 & 1.0 & 1.8 & 4.0 (G) & 1.5 (E) & 3.6 & $_{-2.1}^{+4.1}$ & $_{-3.0}^{+10.3}$
 & 1.5 &$_{-0.9}^{+1.7}$ & 8.2 &$_{-4.7}^{+9.3}$ \\
13 & 1.5 & 1.8 & 4.0 (G) & 1.5 (E) & 2.7 & $_{-1.5}^{+3.0}$ & $_{-2.2}^{+7.6}$
&  1.1 &$_{-0.6}^{+1.3}$ & 6.0 &$_{-3.5}^{+6.8}$ \\
14 & 3.0 & 1.8 & 4.0 (G) & 1.5 (E) & 1.6  & $_{-0.9}^{+1.8}$ & $_{-1.3}^{+4.4}$
 & 0.7 &$_{-0.4}^{+0.7}$ & 3.5 &$_{-2.0}^{+4.0}$ \\
\\
15 & 0.3 & 2.2 & 4.0 (G) & 1.5 (E) & 61.2 & $_{-37.5}^{+75.8}$ & $_{-52.1}^{+190.3}$  &  25.6 &$_{-15.7}^{+31.7}$ & 137.6 &$_{-84.4}^{+170.5}$ \\
16 & 0.7 & 2.2 & 4.0 (G) & 1.5 (E) & 22.1  & $_{-13.4}^{+27.0}$ & $_{-18.7}^{+67.8}$  &  9.2 &$_{-5.6}^{+11.3}$ & 49.7 &$_{-30.2}^{+60.8}$ \\
17 & 1.0 & 2.2 & 4.0 (G) & 1.5 (E) & 14.9 & $_{-9.0}^{+18.0}$ & $_{-12.6}^{+45.2}$ & 6.2 &$_{-3.8}^{+7.5}$ & 33.5 &$_{-20.2}^{+40.5}$ \\
18 & 1.5 & 2.2 & 4.0 (G) & 1.5 (E) & 9.8 & $_{-5.9}^{+11.7}$ & $_{-8.2}^{+29.4}$  & 4.1 &$_{-2.5}^{+4.9}$ & 22.0 &$_{-13.2}^{+26.4}$ \\
19 & 3.0 & 2.2 & 4.0 (G) & 1.5 (E) & 4.7 & $_{-2.8}^{+5.5}$ & $_{-3.9}^{+13.8}$
 &  2.0 &$_{-1.2}^{+2.3}$ & 10.5 &$_{-6.2}^{+12.4}$ \\
\\
20 & 1.0 & 2.5 & 4.0 (G) & 1.5 (E) & 28.3 &$_{-17.5}^{+35.6}$ & $_{-24.2}^{+89.4}$& 11.8 &$_{-7.3}^{+14.9}$ & 63.6 &$_{-39.4}^{+80.0}$ \\
\\
21 & 1.0 & 2.0 & 2.0 (G) & 1.5 (E) & 26.1 &$_{-15.2}^{+29.7}$ & $_{-21.7}^{+74.2}$& 10.9 &$_{-6.3}^{+12.4}$ & 58.6 &$_{-34.1}^{+66.7}$ \\
22 & 1.0 & 2.0 & 3.0 (G) & 1.5 (E) & 12.8 & $_{-7.4}^{+14.6}$ & $_{-10.7}^{+36.4}$  & 5.4 &$_{-3.1}^{+6.1}$ & 28.9 &$_{-16.8}^{+32.8}$ \\
23 & 1.0 & 2.0 & 5.0 (G) & 1.5 (E) & 6.7 & $_{-4.0}^{+7.9}$ & $_{-5.6}^{+19.8}$
 &  2.8 &$_{-1.7}^{+3.3}$ & 15.1 &$_{-8.9}^{+17.8}$ \\
24 & 1.0 & 2.0 & 6.0 (G) & 1.5 (E) & 6.6 & $_{-3.9}^{+7.8}$ & $_{-5.5}^{+19.5}$
 & 2.7 &$_{-1.6}^{+3.3}$ & 14.8 &$_{-8.8}^{+17.5}$ \\
25 & 1.0 & 2.0 & 7.0 (G) & 1.5 (E) & 6.9 & $_{-4.1}^{+8.2}$ & $_{-5.8}^{+20.5}$
 & 2.9 &$_{-1.7}^{+3.4}$ & 15.5 &$_{-9.2}^{+18.4}$ \\
26 & 1.0 & 2.0 & 8.0 (G) & 1.5 (E) & 7.4 & $_{-4.4}^{+8.8}$ & $_{-6.3}^{+22.2}$
 & 3.1 &$_{-1.9}^{+3.7}$ & 16.8 &$_{-10.0}^{+19.9}$ \\
27 & 1.0 & 2.0 & 9.0 (G) & 1.5 (E) & 8.4 & $_{-5.0}^{+10.0}$ & $_{-7.1}^{+25.1}$  & 3.5 &$_{-2.1}^{+4.2}$ & 18.9 &$_{-11.3}^{+22.6}$ \\
 
\enddata
\tablenotetext{1}{Model No.}
\tablenotetext{2}{Minimum luminosity L$_{\rm min}$ in mJy kpc$^{2}$.}
\tablenotetext{3}{Power index of the luminosity function p.}
\tablenotetext{4}{Radial scale length $R_{o}$ in kpc.}
\tablenotetext{5}{Vertical scale height Z$_{o}$ in kpc.}
\tablenotetext{6}{Gaussian (G), and exponential (E) functions for spatial distributions.}
\tablenotetext{7}{Peak value from P(R$_{\rm tot}$).}
\tablenotetext{8}{Confidence level.}
\end{deluxetable}

\clearpage

 \centerline{\psfig{figure=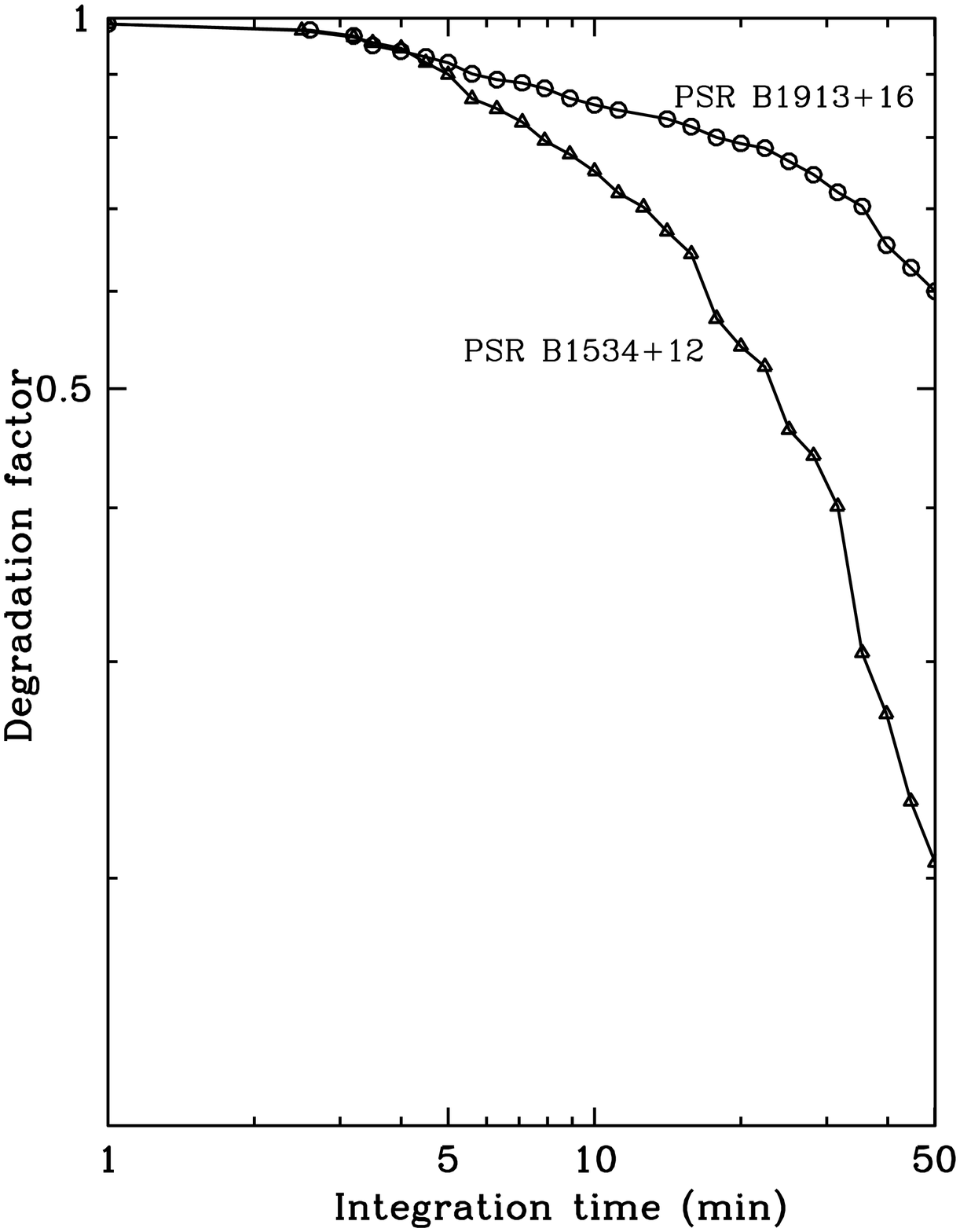,angle=0,width=5.5in}} 
 \figcaption{Average signal-to-noise degradation factor in pulsar search code
versus survey integration time for PSR~B1913+16 and PSR~B1534+12.\label{fig:fvst}}

 \centerline{\psfig{figure=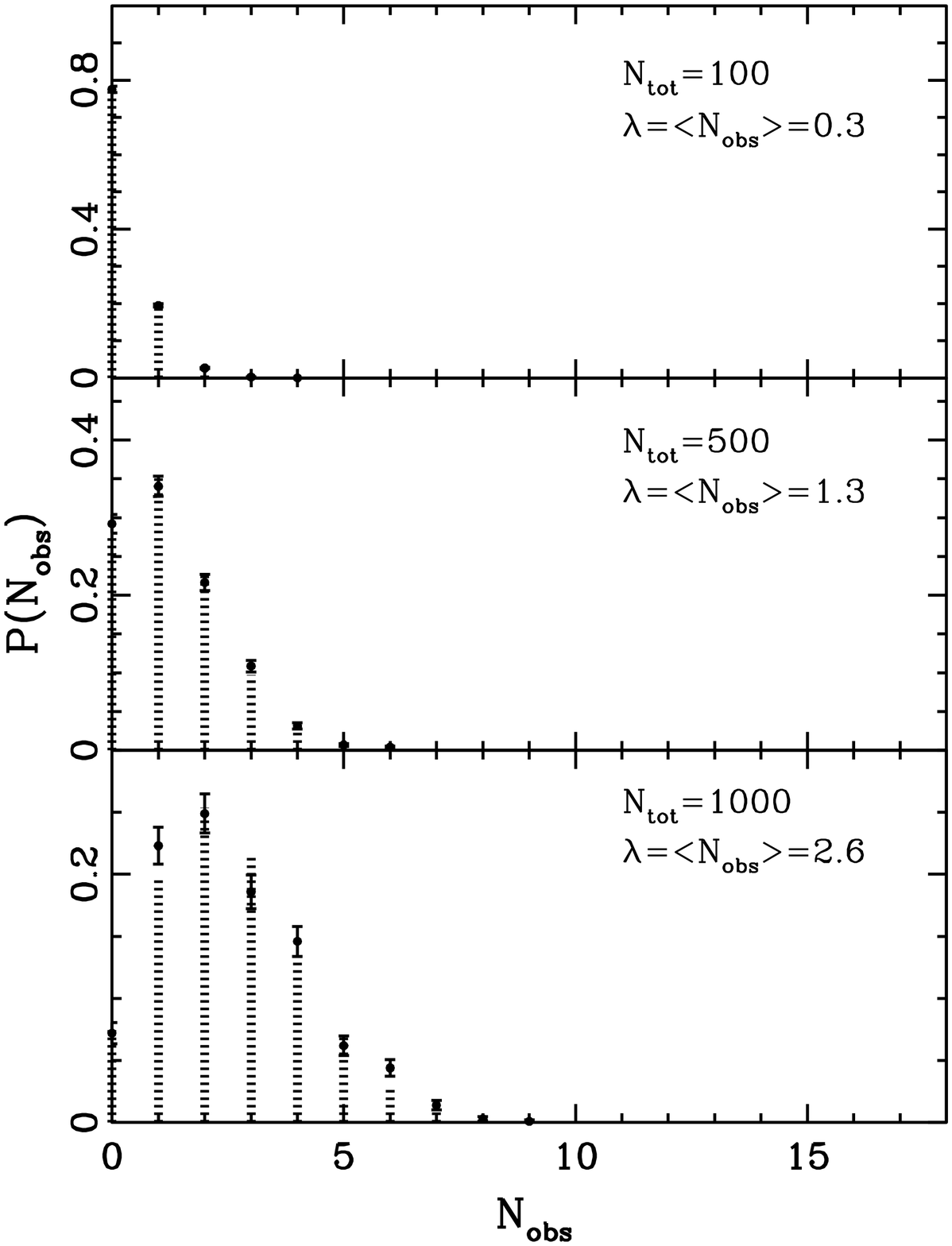,angle=0,width=5.2in}} 
 \figcaption{The Poisson-distribution fits of P(N$_{\rm obs}$) for three
values of the total number $N_{\rm tot}$ of PSR B1913+16-like pulsars in the 
Galaxy (results shown for model 1). 
Points and error bars represent the counts of model samples in our calculation. 
Dotted lines represent the theoretical Poisson distribution.
\label{fig:poisson}}

 \centerline{\psfig{figure=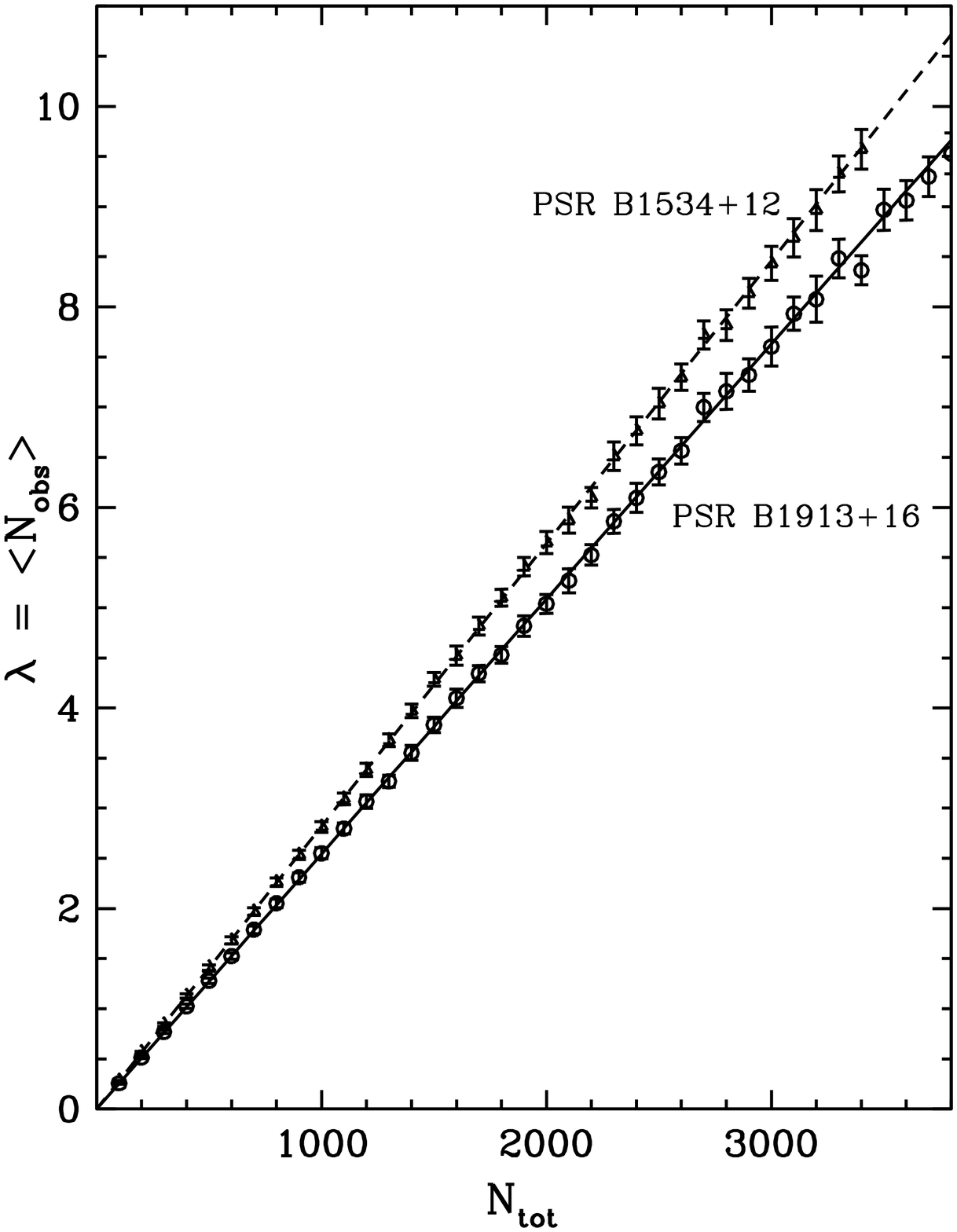,angle=0,width=5.2in}} 
 \figcaption{The linear correlation between $\lambda \equiv <N_{\rm obs}>$ 
and $N_{\rm tot}$ is shown for model 1. Solid and dashed
lines are best-fit lines for PSR~B1913+16-like and PSR~B1534+12-like
populations, respectively. Points and error bars represent the best-fit values 
of $\lambda$ for different values of  $N_{\rm tot}$.
\label{fig:lambda}}

 \centerline{\psfig{figure=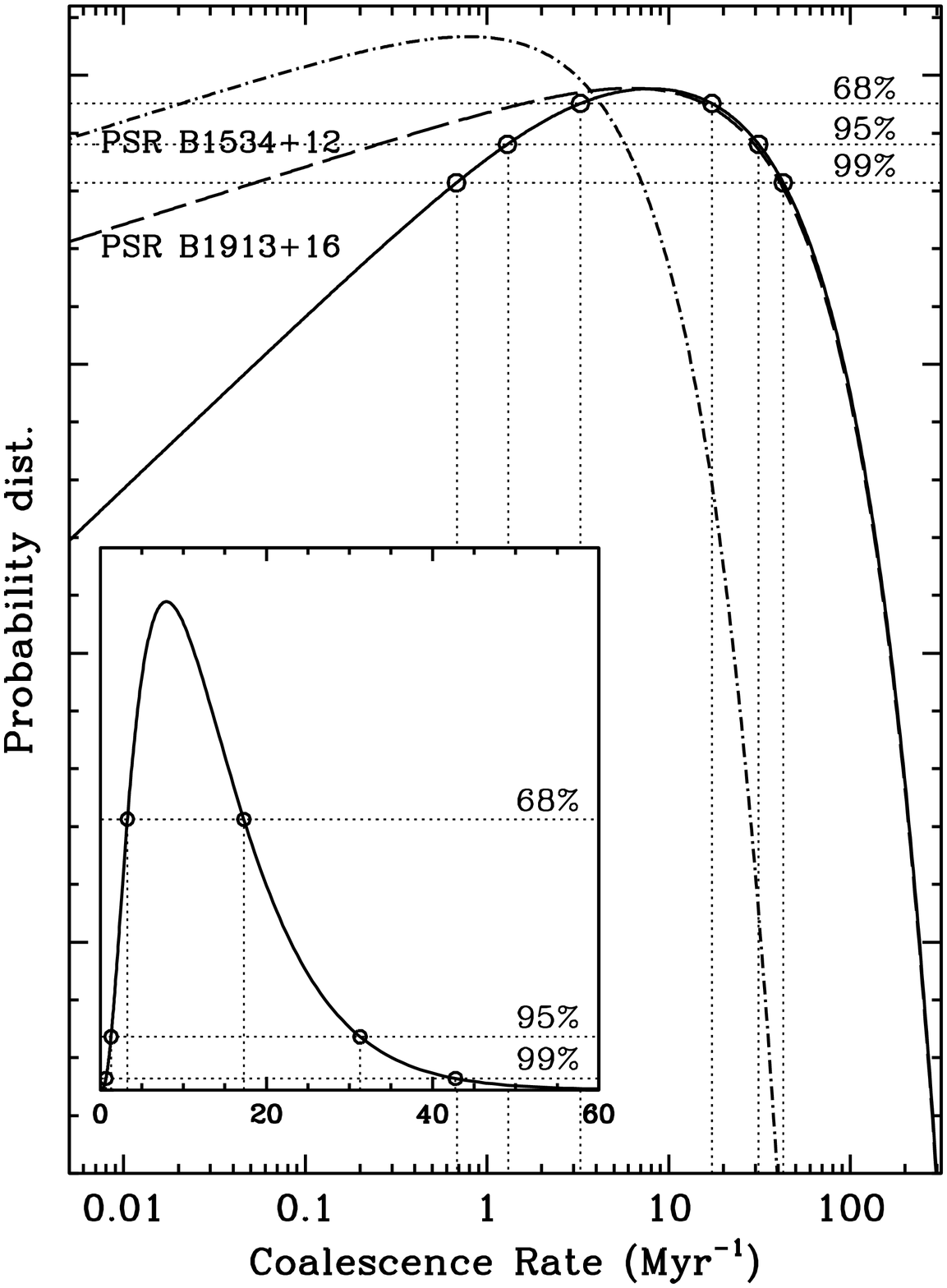,angle=0,width=5.2in}} 
 \figcaption{The probability distribution function of coalescence rates in both a logarithmic and a linear scale (small panel) is shown for model 1. 
The solid line represents $P({\cal R}_{\rm tot})$ and the long and short dashed lines represent $P({\cal R})$ for PSR~B1913+16-like and PSR~B1534+12-like populations, respectively.  We also indicate the confidence levels for $P({\cal R}_{\rm tot})$ by dotted lines.
\label{fig:pdf}}

 \centerline{\psfig{figure=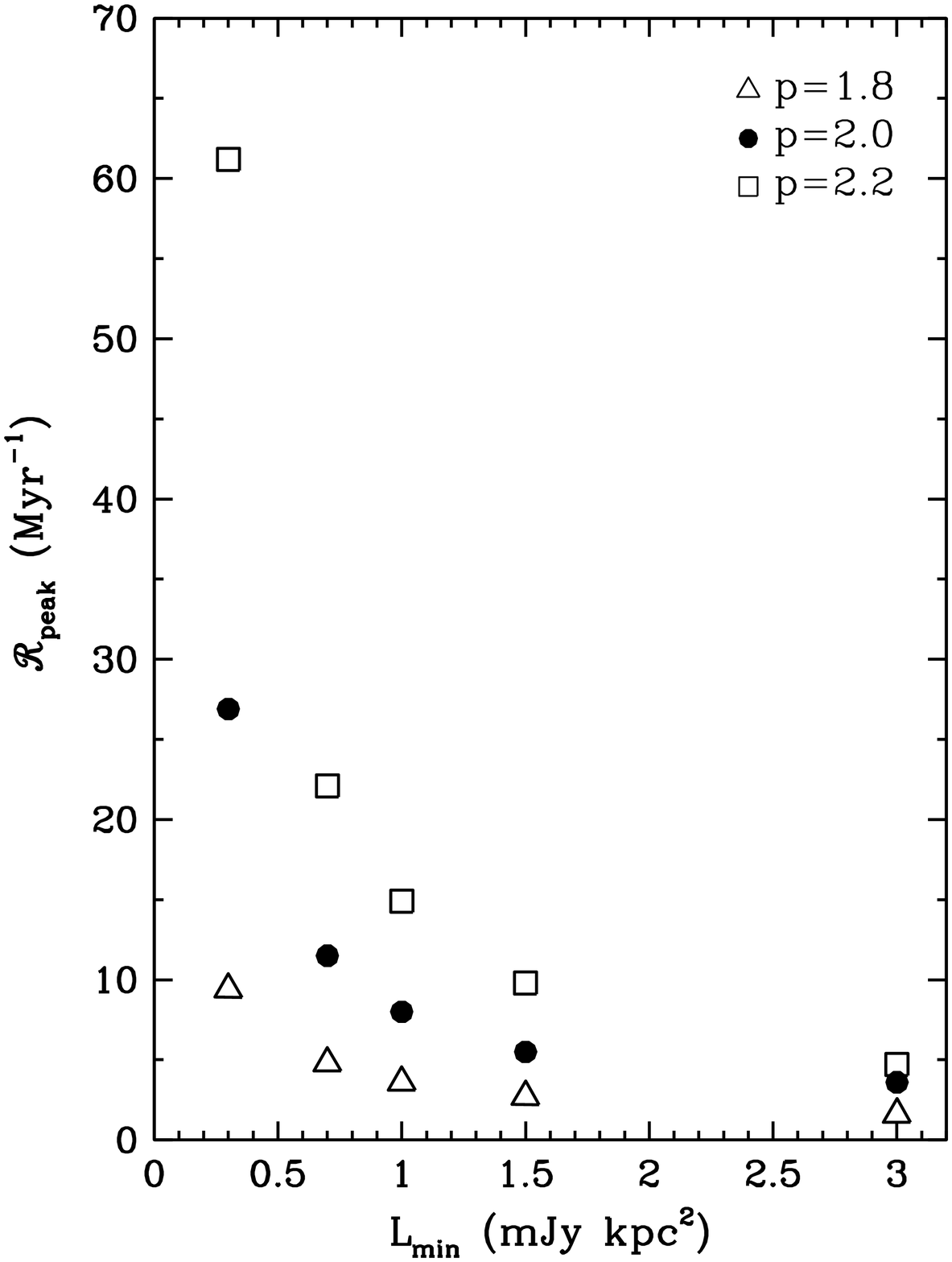,angle=0,width=5.2in}} 
 \figcaption{The correlation between ${\cal R}_{\rm peak}$ and the cut-off luminosity $L_{\rm min}$ with different power indices p of the luminosity distribution function. \label{fig:corr1}} \clearpage

 \centerline{\psfig{figure=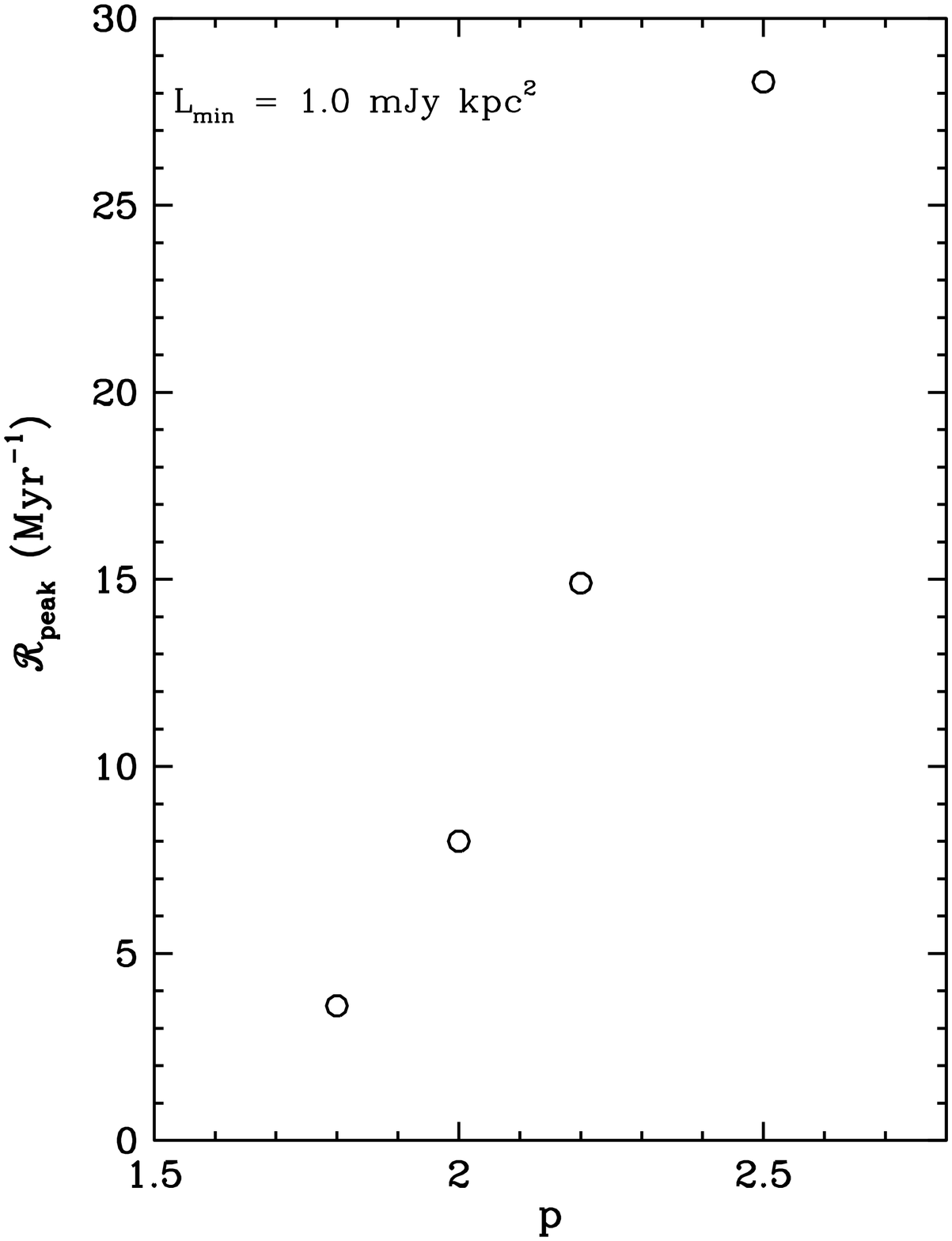,angle=0,width=5.2in}} 
 \figcaption{The correlation between ${\cal R}_{\rm peak}$ and the power index of the luminosity distribution function $p$. \label{fig:corr2}} \clearpage

 \centerline{\psfig{figure=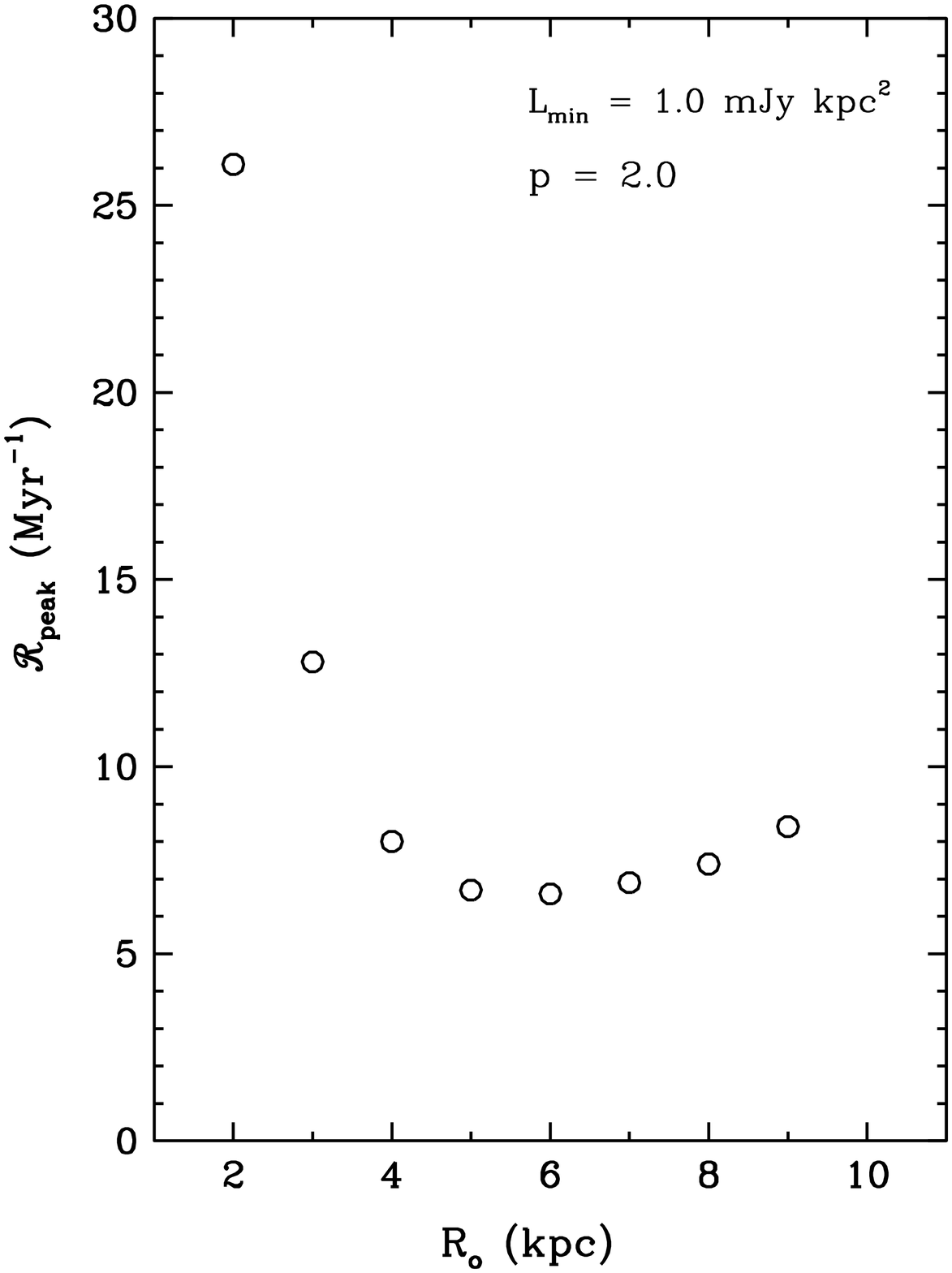,angle=0,width=5.2in}} 
 \figcaption{The correlation between ${\cal R}_{\rm peak}$ and the radial scale length $R_0$. ${\cal R}_{\rm peak}$ is not sensitive to $R_0$ in the range between 4--9 kpc. \label{fig:corr3}} \clearpage

\end{document}